\documentclass[twoside,twocolumn,9pt]{article}
\usepackage{extsizes}
\usepackage[super,sort&compress,comma]{natbib} 
\usepackage[version=3]{mhchem}
\usepackage[left=1.5cm, right=1.5cm, top=1.785cm, bottom=2.0cm]{geometry}
\usepackage{balance}
\usepackage{mathptmx}
\usepackage{sectsty}
\usepackage{graphicx} 
\usepackage{lastpage}
\usepackage[format=plain,justification=justified,singlelinecheck=false,font={stretch=1.125,small,sf},labelfont=bf,labelsep=space]{caption}
\usepackage{float}
\usepackage{fancyhdr}
\usepackage{fnpos}
\usepackage[english]{babel}
\addto{\captionsenglish}{%
  
}
\usepackage{array}
\usepackage{droidsans}
\usepackage{charter}
\usepackage[T1]{fontenc}
\usepackage[usenames,dvipsnames]{xcolor}
\usepackage{setspace}
\usepackage[compact]{titlesec}
\usepackage{hyperref}

\usepackage{epstopdf}

\definecolor{cream}{RGB}{222,217,201}

\begin{document}

\pagestyle{fancy}
\thispagestyle{plain}
\fancypagestyle{plain}{
\renewcommand{\headrulewidth}{0pt}
}

\makeFNbottom
\makeatletter
\renewcommand\LARGE{\@setfontsize\LARGE{15pt}{17}}
\renewcommand\Large{\@setfontsize\Large{12pt}{14}}
\renewcommand\large{\@setfontsize\large{10pt}{12}}
\renewcommand\footnotesize{\@setfontsize\footnotesize{7pt}{10}}
\makeatother

\renewcommand{\thefootnote}{\fnsymbol{footnote}}
\renewcommand\footnoterule{\vspace*{1pt}%
\color{cream}\hrule width 3.5in height 0.4pt \color{black}\vspace*{5pt}} 
\setcounter{secnumdepth}{5}

\makeatletter 
\renewcommand\@biblabel[1]{#1}            
\renewcommand\@makefntext[1]%
{\noindent\makebox[0pt][r]{\@thefnmark\,}#1}
\makeatother 
\renewcommand{\figurename}{\small{Fig.}~}
\sectionfont{\sffamily\Large}
\subsectionfont{\normalsize}
\subsubsectionfont{\bf}
\setstretch{1.125} 
\setlength{\skip\footins}{0.8cm}
\setlength{\footnotesep}{0.25cm}
\setlength{\jot}{10pt}
\titlespacing*{\section}{0pt}{4pt}{4pt}
\titlespacing*{\subsection}{0pt}{15pt}{1pt}

\fancyfoot{}
\fancyfoot[LO,RE]{\vspace{-7.1pt}\includegraphics[height=9pt]{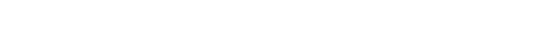}}
\fancyfoot[CO]{\vspace{-7.1pt}\hspace{13.2cm}\includegraphics{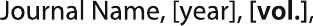}}
\fancyfoot[CE]{\vspace{-7.2pt}\hspace{-14.2cm}\includegraphics{head_foot/RF}}
\fancyfoot[RO]{\footnotesize{\sffamily{1--\pageref{LastPage} ~\textbar  \hspace{2pt}\thepage}}}
\fancyfoot[LE]{\footnotesize{\sffamily{\thepage~\textbar\hspace{3.45cm} 1--\pageref{LastPage}}}}
\fancyhead{}
\renewcommand{\headrulewidth}{0pt} 
\renewcommand{\footrulewidth}{0pt}
\setlength{\arrayrulewidth}{1pt}
\setlength{\columnsep}{6.5mm}
\setlength\bibsep{1pt}

\makeatletter 
\newlength{\figrulesep} 
\setlength{\figrulesep}{0.5\textfloatsep} 

\newcommand{\topfigrule}{\vspace*{-1pt}%
\noindent{\color{cream}\rule[-\figrulesep]{\columnwidth}{1.5pt}} }

\newcommand{\botfigrule}{\vspace*{-2pt}%
\noindent{\color{cream}\rule[\figrulesep]{\columnwidth}{1.5pt}} }

\newcommand{\dblfigrule}{\vspace*{-1pt}%
\noindent{\color{cream}\rule[-\figrulesep]{\textwidth}{1.5pt}} }

\makeatother

\twocolumn[
  \begin{@twocolumnfalse}
{\includegraphics[height=30pt]{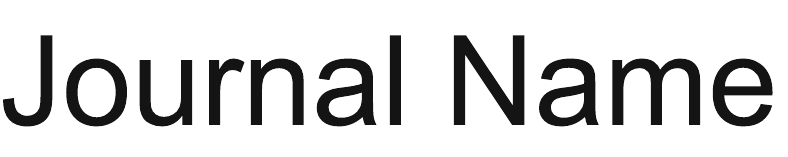}\hfill\raisebox{0pt}[0pt][0pt]{\includegraphics[height=55pt]{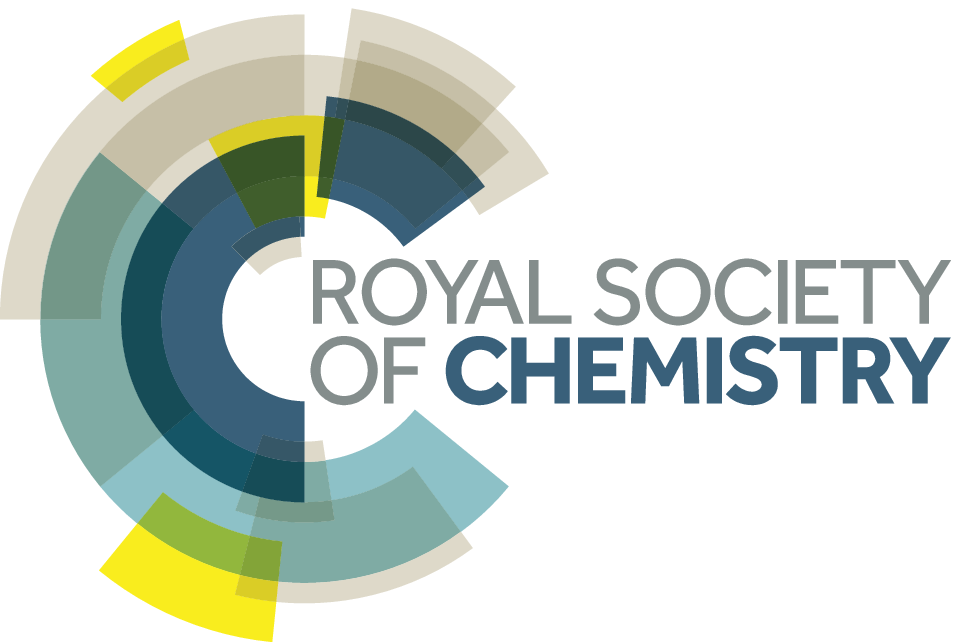}}\\[1ex]
\includegraphics[width=18.5cm]{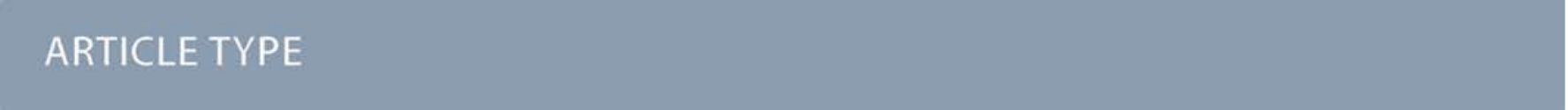}}\par
\vspace{1em}
\sffamily
\begin{tabular}{m{4.5cm} p{13.5cm} }

\includegraphics{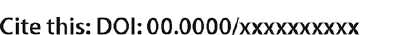} & \noindent\LARGE{\textbf{Na-ion Dynamics in the Solid Solution Na$_{\rm x}$Ca$_{1- \rm x}$Cr$_2$O$_4$ Studied by Muon Spin Rotation and Neutron Diffraction}} \\
\vspace{0.3cm} & \vspace{0.3cm} \\

 & \noindent\large{Elisabetta Nocerino,\textit{$^{a,\dag}$} Ola Kenji Forslund,\textit{$^{b}$} Hiroya Sakurai,\textit{$^{c}$} Nami Matsubara,\textit{$^{a}$} Anton Zubayer,\textit{$^{d}$} Federico Mazza,\textit{$^{e}$} Stephen Cottrell,\textit{$^{f}$} Akihiro Koda,\textit{$^{g}$} Isao Watanabe,\textit{$^{h}$} Akinori Hoshikawa,\textit{$^{i}$} Takashi Saito,\textit{$^{j}$} Jun Sugiyama,\textit{$^{k,l}$} Yasmine Sassa,\textit{$^{b}$} and Martin M\aa{}nsson\textit{$^{a,\ddag}$}} \\

\includegraphics{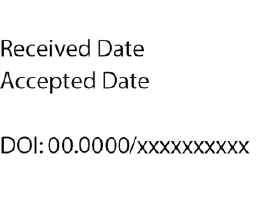} & \noindent\normalsize{In this work we present systematic set of measurements carried out by muon spin rotation/relaxation ($\mu^+$SR) and neutron powder diffraction (NPD) on the solid solution Na$_{\rm x}$Ca$_{1- \rm x}$Cr$_2$O$_4$. This study investigate Na-ion dynamics in the quasi-1D (Q1D) diffusion channels created by the honeycomb-like arrangement of CrO$_6$ octahedra, in the presence of defects introduced by Ca doping. With increasing Ca content, the size of the diffusion channels is enlarged, however, this effect does not enhance the Na ion mobility. Instead the overall diffusivity is hampered by the local defects and the Na hopping probability is lowered. The diffusion mechanism in Na$_{\rm x}$Ca$_{1- \rm x}$Cr$_2$O$_4$ was found to be interstitial and the activation energy as well as diffusion coefficient were determined for all the members of the solid solution.} \\

\end{tabular}

 \end{@twocolumnfalse} \vspace{0.6cm}

  ]

\renewcommand*\rmdefault{bch}\normalfont\upshape
\rmfamily
\section*{}
\vspace{-1cm}


\footnotetext{\textit{$^{a}$~Department of Applied Physics, KTH Royal Institute of Technology, SE-100 44, Stockholm, Sweden}}
\footnotetext{\textit{$^{b}$~Department of Physics, Chalmers University of Technology, SE-412 96 Göteborg, Sweden}}

\footnotetext{\textit{$^{c}$~National Institute for Materials Science, Namiki, Tsukuba, Ibaraki 305-0044, Japan}}
\footnotetext{\textit{$^{d}$~Department of Physics, Chemistry and Biology (IFM), Linköping University, SE-581 83 Linköping, Sweden}}
\footnotetext{\textit{$^{e}$~Insitute of Solid State Physics, TU Wien, Wiedner Haupstraße 8-10, AT-1040 Wien, Austria}}
\footnotetext{\textit{$^{f}$~ISIS Facility, Rutherford Appleton Laboratory, Chilton, Didcot Oxon OX11 0QX, United Kingdom}}
\footnotetext{\textit{$^{g}$~Muon Science Laboratory and Condensed Matter Research Center, Institute of Materials Structure Science, High Energy Accelerator Research Organization (KEK-IMSS), Tsukuba, Ibaraki 305-0801, Japan}}
\footnotetext{\textit{$^{h}$~Muon Science Laboratory, RIKEN, 2-1 Hirosawa, Wako, Saitama 351-0198, Japan}}
\footnotetext{\textit{$^{i}$~Neutron Science Laboratory, Institute of Materials Structure Science, High Energy Accelerator Research Organization, 1-1 Oho, Tsukuba, Ibaraki 305-0801, Japan}}
\footnotetext{\textit{$^{j}$~Institute of Materials Structure Science, High Energy Accelerator Research Organization, 203-1 Shirakata, Tokai, Ibaraki 319-1107, Japan}}
\footnotetext{\textit{$^{k}$~Neutron Science and Technology Center, Comprehensive Research Organization for Science and Society (CROSS), Tokai, Ibaraki 319-1106, Japan}}
\footnotetext{\textit{$^{k}$~Advanced Science Research Center, Japan Atomic Energy Agency, Tokai, Ibaraki 319-1195, Japan}}

\footnotetext{\dag~nocerino@kth.se}

\footnotetext{\ddag~condmat@kth.se}



\section{\label{sec:Intro}Introduction}

A growing population and rapidly developing societies is resulting in an increasing demand for clean energy supply. The harvest, transport, storage and efficient utilization of such energy is one of the grand challenges and fundamental needs of our future sustainable society. In this regard, there is a global drive to change the current energy system and move towards the abandonment of fossil sources and the adoption of renewable ones. Such strive to rebuild the energy system has led to the development of revolutionary materials science and applications, e.g., rechargeable Li-ion batteries \cite{nobel,Etacheri_2011}, hydrogen storage \cite{Niaz_2015,Koppel_2021,Koppel_2022}, photovoltaics \cite{Nayak_2019}, and carbon capture \cite{Wilberforce_2021}. Such technologies are some of the most efficient ways to promote a shift towards sustainable development and away from fossil fuels. There are though still many challenges that needs to be addressed. For instance, in the case of Li-ion batteries, there are issues regarding scarcity and uneven geographical distribution of the resources required for the production of Li-batteries (Li, Co, Cu, Ni), as well as the high environmental impact of their extraction and the consequential high (constantly growing) manufacturing costs. As a result, such technology might soon become both environmentally as well as economically non viable \cite{kavanagh2018global}. This fact raises the question whether the environmental advantages of Li-batteries are canceled by the non-sustainability of their production \cite{Alexander_2020}.
On one hand, part of the scientific community is trying to face the problem with new strategies for lithium mining and recycling \cite{agusdinata2018socio}, on the other hand more sustainable alternatives are being investigated. For the latter, replacing Li with similar alkali ion like Na \cite{Hwang_2017} or K \cite{Kanyolo_2021} is currently a very active field of research. 
The Na counterparts of Li-ion batteries experienced during the past decades a steadily increasing interest, from academic as well as industrial researchers, due to the undeniable advantages of Na-based batteries \cite{peters2019exploring}. Beyond the fact that Na is one of the most abundant elements on Earth's crust as well as sea water, and therefore much more accessible than Li, the main difference between the two kinds of batteries lies in the nature of the cathode material \cite{pillot2017avicenne}. While the preparation costs and procedures are comparable for both battery types, controversial elements like cobalt, required in Li-batteries, are not necessary in the production of Na-batteries. As a result, a dramatic reduction of the manufacturing costs as well as improvement of the socio-environmental impact of this technology is achieved. Researchers in this framework are encouraged to study advanced cathode (and anode) materials, in order to gain a deeper understanding of their fundamental properties. Here, one of the key aspects is to understand the link between the crystal structure and ion transport at the atomic level. The aim is to provide the know-how to systematically tailor high-capacity, reversible electrodes for Na-ion batteries.

The focus of this work is the detailed study of how the crucial mechanism for Na-ion transport is linked to subtle changes in the crystal structure within the quasi-1D NaCr$_2$O$_4$ compound. The influence from substituted Ca ions on the Na-ion diffusive behavior in the solid solution Na$_{\rm x}$Ca$_{1- \rm x}$Cr$_2$O$_4$ is systematically investigated. Such system is utilized as a model system for defects in current and future low-dimensional battery materials, e.g. the well established cathode material LiFePO$_4$ \cite{Sugiyama2011, Benedek_2019, Benedek_2020}.


\begin{figure*}[ht]
  \begin{center}
    \includegraphics[keepaspectratio=true,width=175 mm]{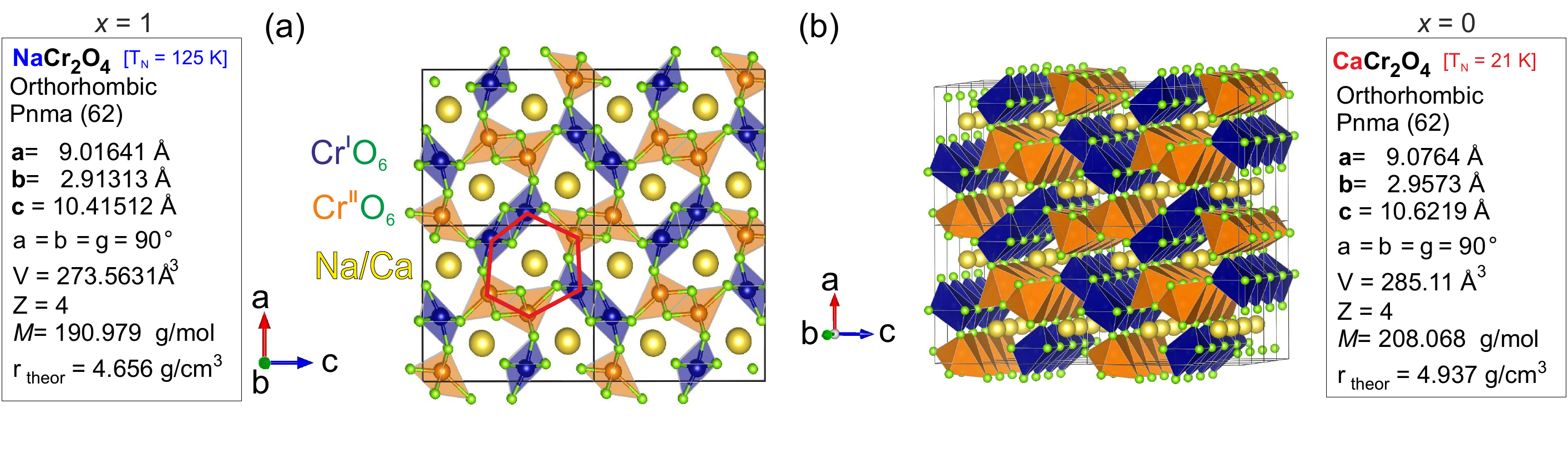}
  \end{center}
  \caption{Crystal structure of the Na$_{\rm x}$Ca$_{1- \rm x}$Cr$_2$O$_4$ family and cell parameters for the two extremes of the solid solution NaCr$_2$O$_4$ ($x=1$) and CaCr$_2$O$_4$ ($x=0$). Figure (a) shows the structure seen from the $ac-$plane, where the hexagonal 1D channels hosting the Na$^+$ (Ca$^{2+}$) ions is emphasize in red. (b) The octahedral chains and 1D Na/Ca channels aligned along the $b-$axis are clearly visible.}
  \label{struct}
\end{figure*}

NaCr$_2$O$_4$ (formally NaCr$^{3+}$Cr$^{4+}$O$_4$) is a spinel-ferrite \cite{muller} (calcium ferrite-type), quasi-1D (Q1D) novel transition metal oxide \cite{sakurai}, posing Cr$^{3+}$/Cr$^{4+}$ mixed-valence state. Such an oxidation state for the Cr ion is very unusual, and it can only be stabilized under extreme pressure conditions (synthesis). There are very few realizations of this phenomenon in actual samples: beyond NaCr$_2$O$_4$, another material that exhibits Cr$^{3+}$/Cr$^{4+}$ mixed-valence state is K$_2$Cr$_8$O$_18$ \cite{forslund2018mu+}. This rare condition is responsible for unconventional low temperature microscopic properties \cite{kolodiazhnyi2013electronic} that make NaCr$_2$O$_4$ very interesting also in a fundamental research perspective.
The compound crystallizes in the orthorhombic space group 62 (\textit{Pnma}), in which Cr cations occupy two distinct crystallographic positions (labelled as Cr$^I$ and Cr$^{II}$ in Fig.~\ref{struct}) surrounded by octahedrally coordinated oxygen atoms. The CrO$_6$ octahedra are in turn arranged in double zig-zag chains, by sharing one edge along the $b-$axis. The Na$^+$ ions are located in the hexagonal one-dimensional channels designed by the interconnections amongst the different chains [Fig.~\ref{struct}(a)], which are believed to be privileged directions for ion diffusion.
Na$_{\rm x}$Ca$_{1- \rm x}$Cr$_2$O$_4$ is obtained as a solid solution between the iso-structural compounds $\beta$-CaCr$_2$O$_4$ - NaCr$_2$O$_4$. The evolution of the electronic and, partially, the spin structure for the family Na$_{\rm x}$Ca$_{1- \rm x}$Cr$_2$O$_4$ was studied as a function of the Na content by X-ray absorption spectroscopy (XAS)\cite{taguchi2017unusual}, neutron diffraction (ND) and bulk $\mu^+$SR \cite{jun}. The Na$^+$ substitution for Ca$^{2+}$ introduces holes in the electronic state, leading to the partial oxidation of Cr$^{3+}$ to Cr$^{4+}$, and induces a change in the magnetic ordering of the Cr moment from incommensurate anti-ferromagnetic (IC-AF) structure in CaCr$_2$O$_4$ to commensurate anti-ferromagnetic (C-AF) structure in NaCr$_2$O$_4$. As a result, hole doping-induced charge frustration and magnetic interaction-induced geometrical frustration of the lattice occur in NaCr$_2$O$_4$ \cite{taguchi2017unusual,jun}.

In this work a systematic doping-dependent study on Na$_{\rm x}$Ca$_{1- \rm x}$Cr$_2$O$_4$ with [x = 0.3, 0.5, 0.7, 0.85, 0.90, 0.95, 1] is presented. The investigation is carried out using muon spin rotation/relaxtaion ($\mu^+$SR) and neutron powder diffraction (NPD), to show how the size and the ionic content of the 1D CrO$_6$ diffusion channels affects the kinetics of Na ions. Moreover, if the Ca ions are regarded as "defects", this study provides a description of phenomena occurring in low dimensional battery materials affected by defects. The element Ca is especially suitable for this kind of investigation since it has zero nuclear magnetic moment, which makes it imperceptible to the muons. This fact ensures that the dynamic behavior observed in the muon signal comes from the Na-ions. The results obtained evidence a clear trend for the diffusion becoming more and more hampered as the Ca content increases.

\section{\label{sec:exp}Experimental Setup}

Polycrystalline samples of Na$_{\rm x}$Ca$_{1- \rm x}$Cr$_2$O$_4$ were prepared from stoichiometric mixtures of CaO, NaCrO$_2$, Cr$_2$O$_3$, and CrO$_3$ at 1300$^{\circ}$C under a pressure of 6 GPa, while the NaCr$_2$O$_4$ was prepared under a pressure of 7 GPa.
Further information regarding the sample synthesis can be found in reference \cite{sakurai}. All the samples were synthesized at the National Institute for Material Science (NIMS) in Tsukuba Japan. From powder X-ray diffraction (XRD) they were proven to be single phase, with a CaFe$_2$O$_4$-type $\textit{Pnma}$ structure.

The $\mu^+$SR spectra were acquired at the muon spectrometers EMU \cite{giblin2014optimising} and RIKEN-Ral \cite{nagamine1994construction}, at the ISIS Neutron and Muon source \cite{houck2000isis} (United Kindom), and S1 \cite{kojima2018development} at the J-PARC research facility \cite{nagamiya2012introduction} (Japan). For the muon measurements, $\sim1.5$~g of sample in powder were pressed into a pellet under a pressure of about 1.9 tons. The pellet was sealed in a 23.5~mm diameter Ti cell, with Ti screws, a Ti window of $50\mu$m thickness and a gold O-ring for sealing. The sample was then mounted on a closed cycle refrigerator to reach temperatures from 50 K to 600 K.


The neutron powder diffraction (NPD) patterns were collected at the time-of-flight (ToF) powder diffractometers iMATERIA \cite{ishigaki2009ibaraki} and SPICA \cite{yonemura2014development} at J-PARC. The neutron diffraction measurements were performed on powder samples ($\sim$ 0.72 g) mounted into cylindrical vanadium cells with diameters 6 mm (for SPICA) and 5 mm (for iMATERIA). The cell was closed with an aluminium cap, aluminium screws and indium sealing. The cell was mounted on a closed cycle refrigerator to reach temperatures between 2 K and 300 K. While the low temperature properties of the solid solution Na$_{\rm x}$Ca$_{1- \rm x}$Cr$_2$O$_4$ will be discussed elsewhere, the current work will be focused on the room temperature results.

The software VESTA \cite{momma} was used for crystal structure visualization, {\sf MATLAB} \cite{mat} and {\sf IGOR Pro} (Wavemetrics, Lake Oswego, OR, USA) \cite{igor} for parameter plotting and fitting. The software {\sf \textit{musrfit}} \cite{suter2012musrfit} was used for the fit of the $\mu^+$SR data, while {\sf Fullprof} \cite{rodriguez2001fullprof} for the neutron diffraction data analysis. The Bilbao Crystallographic Server has been often consulted during the preparation of this paper \cite{aroyo2011crystallography, aroyo2006bilbao, aroyo2006bilbao2}.

\section{\label{sec:results}Results and Discussion}
In the following section the experimental results with the related data analysis are collected.

\subsection{\label{neut}Structural evolution studied by Neutron Diffraction}

In order to study the structural evolution of the solid solution Na$_{\rm x}$Ca$_{1- \rm x}$Cr$_2$O$_4$, room temperature neutron diffraction patterns were collected for the compositions with [$x = 0.5,~0.7,~0.9$] at the instrument SPICA (high angle detector bank, resolution $\Delta d / d$ = 0.12$\%$) and for [$x=1$] at the instrument iMATERIA (backward detector bank, resolution $\Delta d / d$ = 0.16$\%$). The data with their corresponding Rietveld refinement are displayed in Fig. \ref{neut}. 
The observed profiles are in agreement with the calculated models. The goodness of each model is underlined by the values of the reliability R-factors (reported in Fig. \ref{neut}), none of them exceeding a few percent. The values of the $\chi^2$, slightly larger than the ideal value 1, can be justified by the very high resolution of the diffractometers.

\begin{figure}[h!]
  \begin{center}
    \includegraphics[keepaspectratio=true,width=68 mm]{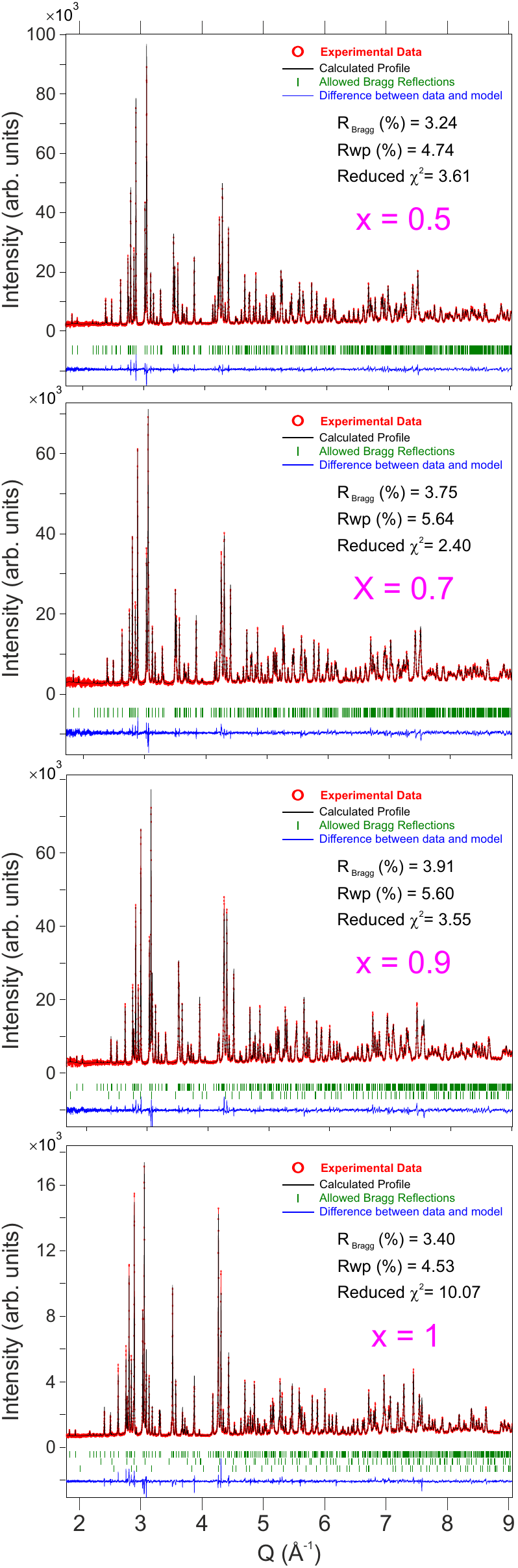}
  \end{center}
  \caption{Neutron powder diffraction patterns at $T=300$~K for different Na contents ($x$) in the solid solution Na$_{\rm x}$Ca$_{1- \rm x}$Cr$_2$O$_4$ with the respective values of the reliability factors and $\chi^2$. See also Table~\ref{cell} and Table~\ref{dist} for detailed fitting results.}
  \label{neut}
\end{figure}

The Bragg peaks were initially indexed as nuclear peaks for Na$_{\rm x}$Ca$_{1- \rm x}$Cr$_2$O$_4$, using the aforementioned $\textit{Pnma}$ space group, via de Le Bail method. Tiny amounts of Cr$_2$O$_3$ and CrO$_2$ impurities were also present in the samples, and their Bragg peaks could be easily separated and indexed using the space groups 167 and 136 respectively. The atomic positions were extracted by Rietveld refinement. Here the positions of Ca and Na were refined together as they occupy the same crystallographic site. The data were corrected for absorption for a cylindrical sample and the chosen peak shape function is a pseudo-Voigt convoluted with a pair of back-to-back exponentials. The background was fitted with a linear interpolation of manually added points. Isotropic atomic displacement parameters have been refined. The resulting cell parameters for the main phase are reported in Table~\ref{cell}. As expected, the size of the unit cell increases isotropically as the Ca doping increases with the lattice being less and less frustrated \cite{kolodiazhnyi2013electronic}.

\begin{table}[ht]
\caption{\label{cell}
Cell parameters of Na$_{\rm x}$Ca$_{1- \rm x}$Cr$_2$O$_4$ extracted from NPD measurements as a function of the Na content ($x$) at $T=300$~K. Space group for all compositions are \textit{Pnma}.
}

\renewcommand{\arraystretch}{1.25}
\begin{tabular}{ccccc}
 \hline
 Na content ($x$) &a (Å) &b (Å) &c (Å) &\\
  \hline
   0.5 &9.0418(1)&2.9297(1)&10.5640(2)&\\
 
   0.7 &9.0248(9)&2.9198(8)&10.5199(1)&\\

   0.9 &9.0119(1)&2.9139(1)&10.4568(1)& \\
 
   1 &9.0154(1)&2.9128(1)&10.4138(9)&  \\
 \hline
\end{tabular}
\end{table}

Table~\ref{dist} lists the values of the atomic distances between chromium and oxygen $d_{\rm Cr1(2)-O}$ (the two crystallographic different sites Cr1 and Cr2 have comparable distances from O) and between sodium and oxygen $d_{\rm Na-O}$. As the Na concentration decreases, $d_{\rm Cr1(2)-O}$ increases. This is due to the gradual increase of charge separation for Cr atoms going from a mixed valence 3.5+ in the pure Na compound ($x=1$) to the valence 3+ in the doped compound with $x=0.5$, which results in the occurrence of chemical pressure effect. The distance $d_{\rm Na-O}$ instead shows an opposite trend since it decreases with lower Na content. A bigger distance between Na and O implies weaker bonds between the two, which might be one of the factors that promotes Na-ion mobility.

\begin{table}[ht]
\caption{\label{dist}
Comparison between atomic distances in the four samples, as extracted from the Rietveld refinement of the NPD data at $T=300$~K.
}
\renewcommand{\arraystretch}{1.25}
\begin{tabular}{ccccc}
 \hline
 Na content ($x$) &$d_{\rm Cr1-O}$ (Å)~$\approx$ $d_{\rm Cr2-O}$ (Å)& &$d_{\rm Na-O}$ (Å) &\\
 \hline
   0.5 &1.99& &2.49&\\
 
   0.7 &1.98& &2.51&\\
 
   0.9 &1.96& &2.53& \\
 
   1 &1.95& &2.54&  \\
\hline
\end{tabular}
\end{table}

\begin{figure}[ht]
  \begin{center}
    \includegraphics[keepaspectratio=true,width=77 mm]{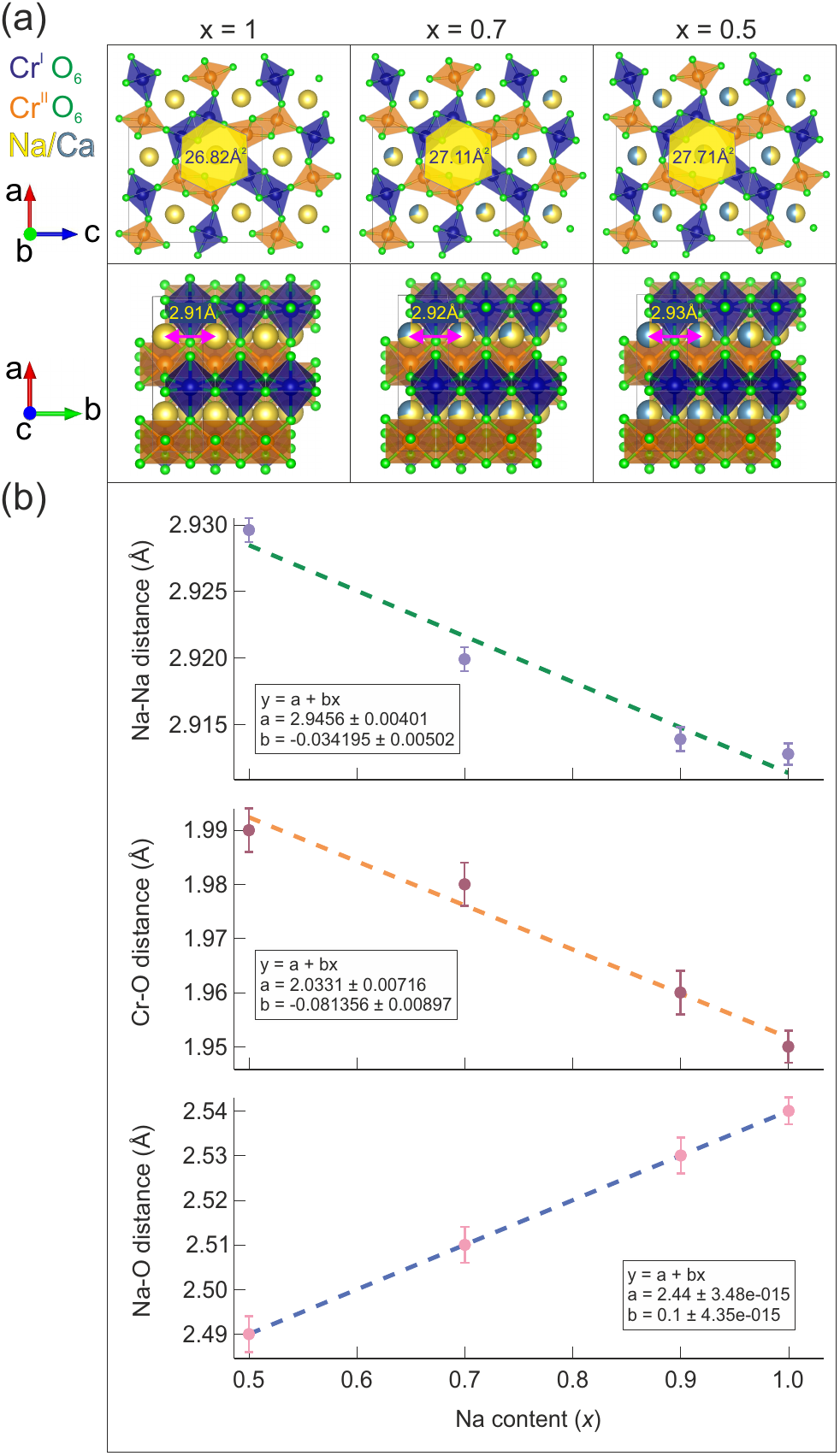}
  \end{center}
  \caption{(a) Evolution of the 1D Na/Ca channels as a function of the Na content $x$. The area of the channel's cross section is highlighted in yellow. Also shown is the Na-Na(ca) distance. (b) distances between consecutive Na/Ca sites, Cr-O sites and Na-O sites as a function of the Na content $x$. The dashed lines are linear fits to the data.}
  \label{chan}
\end{figure}

A model for the size of the 1D diffusion channels as a function of the Na content is displayed in Fig.~\ref{chan}(a). The size of the channels is estimated by geometrical considerations. It is easy to see how the channels become smaller, as an effect of the reduced unit cell volume. The atomic distances as a function of the Na content ($x$) are also plotted in Fig.~\ref{chan}(b) and the distances manifest a clear linear trend as a function of $x$.


A recent study by Byles and Pomerantseva on tunnel structured manganese oxides \cite{byles2021effect}, carried out by galvanostatic intermittent titration and electronic conductivity measurements, reports an interesting relationship between the size of the structural tunnel (or 1D diffusion channel) and the diffusive behavior of the charge carrying ion. In particular, they show a comparison among the rate performances of Li-ion and Na-ion battery materials (LIB and NIB respectively), with channels of different sizes built by edge sharing MnO$_6$ octahedra. The study shows that the material with the largest channels provided the best performance for the LIBs but not for the NIBs, in which smaller tunnels with more stable and well defined Na sites characterized the best performing material.
A similar behavior can be recognized in Na$_{\rm x}$Ca$_{1- \rm x}$Cr$_2$O$_4$. From our current $\mu^+$SR study (see next section), the mobility rate of Na-ions actually increases with the Na content. This means that Na-ion diffusion is enhanced with a narrower diffusion channel.

\subsection{\label{musr}Na-ion Diffusion studied by $\mu^+$SR}
\begin{figure}[ht]
  \begin{center}
    \includegraphics[keepaspectratio=true,width=65 mm]{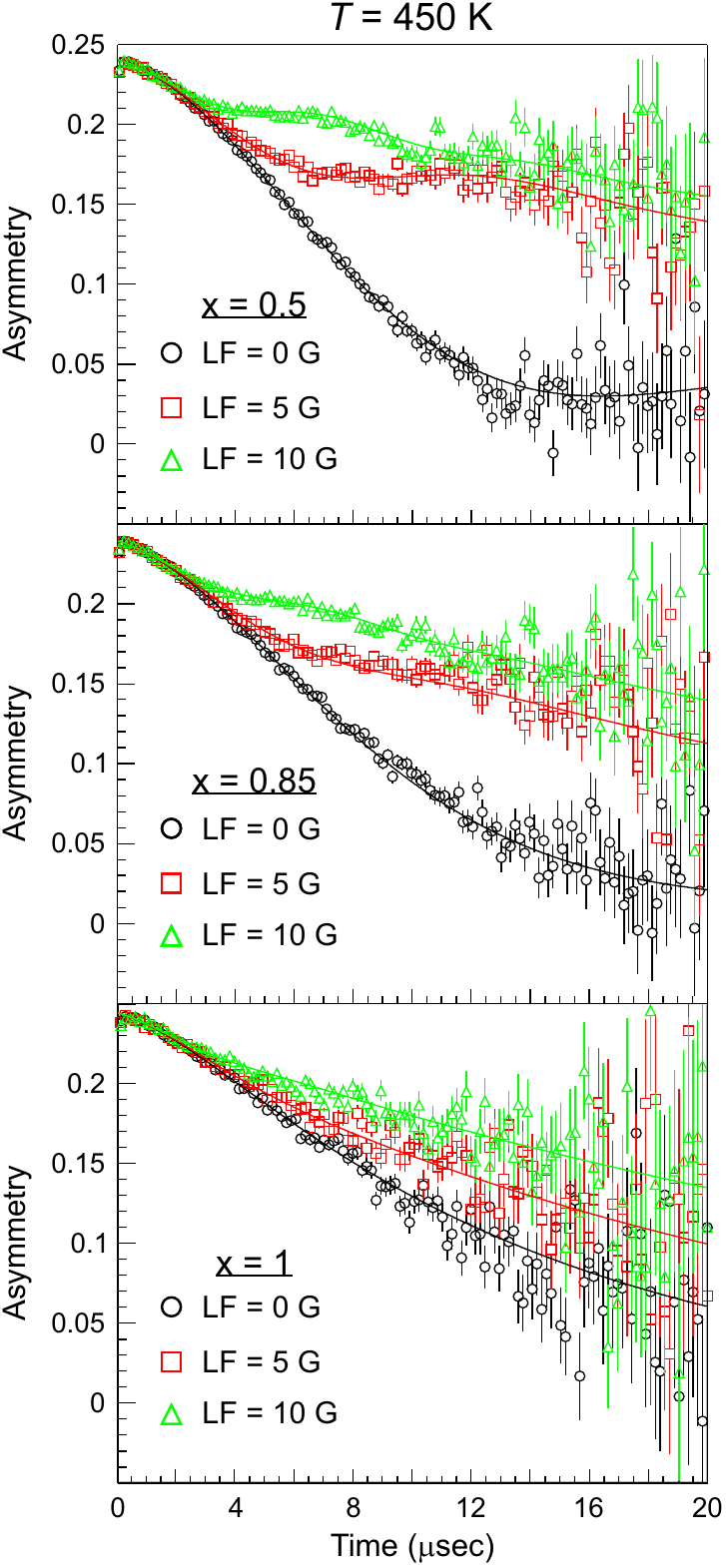}
  \end{center}
  \caption{$\mu^+$SR time spectra collected at $T=450$~K for different Na content ($x$) and values of the applied longitudinal magnetic field (LF). The solid lines are best fits to the data using Eq.~\ref{3} with a global fitting protocol for each temperature.}
  \label{muon}
\end{figure}

A regular and systematic use of the $\mu^+$SR technique for studies of ion dynamics only began about a decade ago \cite{maansson2013muon,Sugiyama2009}. However, since then there is a strong and steady increase of such reports in the published literature, covering \textit{ex-situ} material studies for batteries \cite{Sugiyama2011,Benedek_2020,Ma_2021}, H-storage \cite{Sugiyama_2010}, and photvoltaics \cite{Ferdani_2019}, as well as more recent \textit{in-situ/-operando} investigations \cite{Sugiyama_2019,McClelland_2021,Ohishi_2022}. The principle behind this rather unique method lays in the ability of the particle probe (anti-muons $\mu^+$) to detect fluctuating magnetic moments, originated from ion diffusion in solids \cite{sugiyama2013ion}. As the spin polarized muon beam is implanted into the sample, the muon spin precesses according to the local magnetic environment. In particular, considering the most general case, muons can sense the hyperfine fields due to fluctuating electronic spins, the static nuclear dipolar fields from the atoms in the lattice, and also the modulated nuclear dipolar fields due to fluctuating nuclear spins coupled to the fluctuating electronic spins \cite{hayano1978observation}. For the present study we will assume that the latter interaction term does not contribute to the detected muon signal. The application of an external magnetic field of weak intensity (comparable to the modulated nuclear dipolar field $\approx$ 10~G), whose flux lines are parallel to the initial direction of the polarized muon spin (so-called longitudinal field LF), allows us to partly decouple the contribution from nuclear and electronic spins \cite{matsuzaki1987critical}. This protocol makes it possible to distinguish between the internal magnetic fields arising from nuclear and electronic contributions, as the muon relaxation will be different in the two cases. In this way, it ensures a robust determination of ion-diffusion related changes in the nuclear dipolar field.
The $\mu^+$SR spectra of Na$_{\rm x}$Ca$_{1- \rm x}$Cr$_2$O$_4$ at $T=450$~K for selected compositions are displayed in Fig.~\ref{muon}. Each sample was measured in the same conditions: under zero field (ZF), as well as under 5~G and 10~G decoupling LF (and also one weak transverse field calibration, not shown).
As seen in Fig.~\ref{muon}, for the sample $x=0.5$ the ZF muon spin relaxation display a typical Kubo-Toyabe (KT) function for a isotropically distributed nuclear dipolar field.
The presence of the decoupling LF causes a reduction of the signal's relaxation rate as the muon spins are 'locked' in their initial direction by the applied field.

\begin{figure}[ht]
  \begin{center}
    \includegraphics[keepaspectratio=true,width=75 mm]{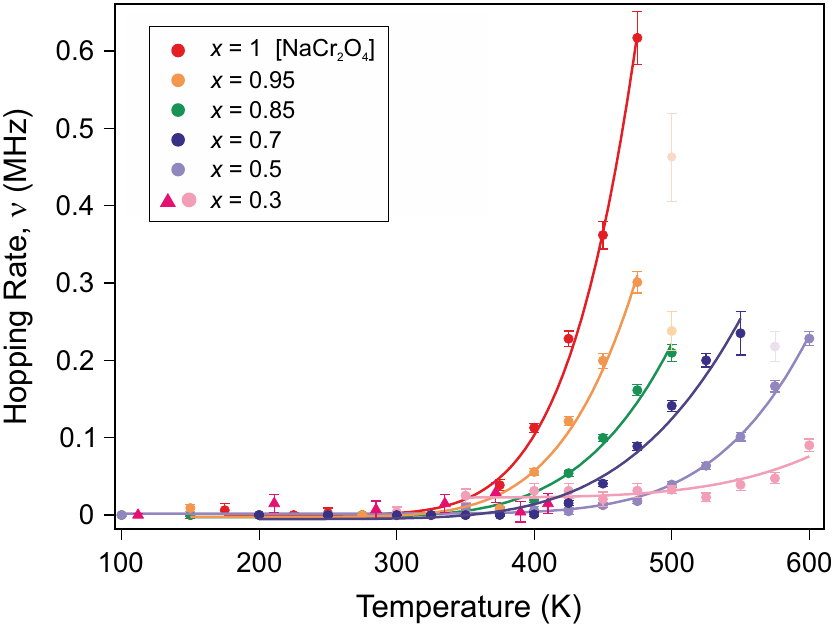}
  \end{center}
  \caption{Plot of the Na-ion hopping rate $\nu$ as a function of temperature for different Na contents ($x$). A clear thermally activated exponential increase is observed, which is typical for a diffusion process. The solid lines are fits to the data using an Arrhenius function (Eq.~\ref{arr}).}
  \label{hop}
\end{figure}

Comparing the three plots in Fig.~\ref{muon} it is evident that, as the Na content increases, the muon relaxation rate is reduced (also for the ZF case) as the tail of the signal in the long time domain drifts upwards. This behavior can be explained by observing the temperature and composition dependent trends extracted from careful fitting of the $\mu^+$SR time spectra. The fit function chosen for the entire Ca$_{1-x}$Na$_{x}$Cr$_2$O$_4$ family is the following:

\begin{eqnarray}
 A_0 \, P_{\rm LF}(t) &=&
A_{\rm KT}G_{\rm DGKT} (\sigma, \nu, t, H_{\rm LF}) \cdot{}e^{(-\lambda_{\rm KT} t)}
\cr
 &+& A_{\rm BG}.
\label{3}
\end{eqnarray}
The function is constituted by the sum of a dynamic gaussian Kubo-Toyabe (KT) relaxation function G$_{\rm DGKT}$, multiplied by an exponential relaxation function, plus a small background signal from the muons stopping in the sample holder A$_{\rm BG}$. Here, A$_0$ is the initial asymmetry of the muon decay (maximum value); A$_{\rm KT}$ and A$_{\rm BG}$ are the asymmetries of the KT function and of the background respectively. These quantities provide an estimate of the volume fractions of the muons implanted in the sample, whose behavior is described by the related relaxation function. The dynamic gaussian KT describes the depolarization of the muon spin in a fluctuating nuclear dipolar field, characterized by a gaussian distribution. G$_{\rm DGKT}$ is a function of several parameters: $\sigma$ is related to the width of the internal field distribution $\Delta$ by the relation $\sigma = \gamma_{\mu} \sqrt{\Delta}$ (here $\gamma_{\mu}$ is he muon gyromagnetic ratio); $\nu$ is the fluctuation rate of the field at the muon sites (which in this case translates into the Na-ion hopping rate); $t$ is the time and $H_{\rm LF}$ is the externally applied longitudinal field. The exponential decay rate $\lambda_{\rm KT}$ is due to the rapidly fluctuating electronic moments of the transition metal Cr atoms in the paramagnetic state. The exponential relaxation rate, was found to be virtually independent of temperature and composition. In the final fits it was therefore kept fixed to its room temperature value $\lambda_{\rm KT} = (0.02 \pm 0.005)~\mu$s$^{-1}$. The condition $\nu=0$ and $H_{\rm LF}$~=~0 corresponds to the static ZF case, in which the KT function describes the depolarization of the muon spin in a static nuclear dipolar field, arising from randomly oriented nuclear spins. Following upon these assumptions, the ZF and the two LF muon spectra for each temperature have been fitted simultaneously while keeping $\Delta$ and $\nu$ as common variables. This results in the reliable determination of both the static ($\Delta$), as well as dynamic ($\nu$), parameters. Figure \ref{hop} displays the temperature dependence of the hopping rate for each sample in the solid solution.

\begin{figure}[ht]
  \begin{center}
    \includegraphics[keepaspectratio=true,width=75 mm]{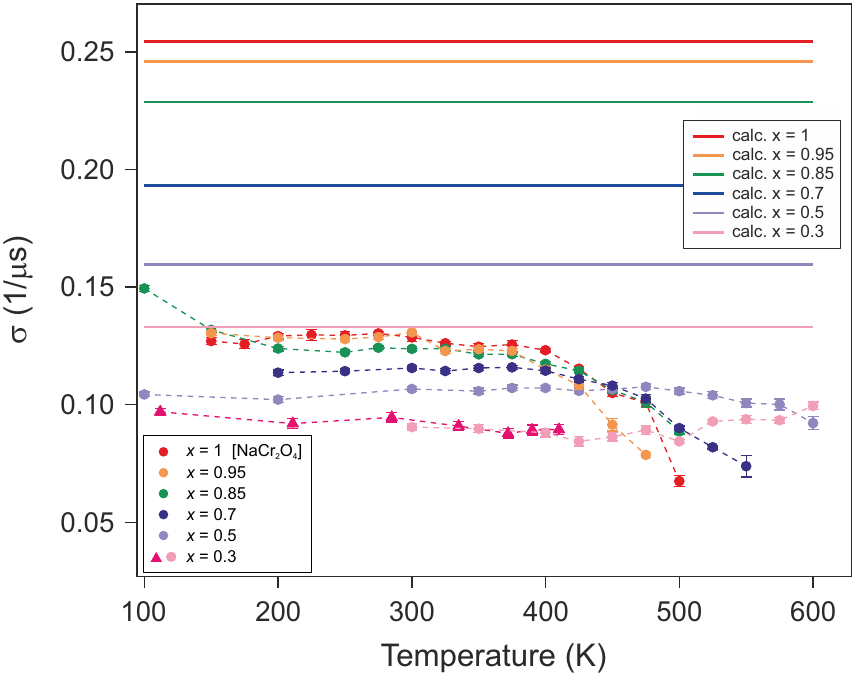}
  \end{center}
  \caption{Comparison between the measured (dots) and calculated (solid lines) $\sigma$. The discrepancy between the measured and calculated values is due to the fact that the calculated $\sigma$ is evaluated in T $\rightarrow$ 0. The overall trend is, however, fully consistent.}
  \label{delta}
\end{figure}

\begin{figure}[ht]
  \begin{center}
    \includegraphics[keepaspectratio=true,width=75 mm]{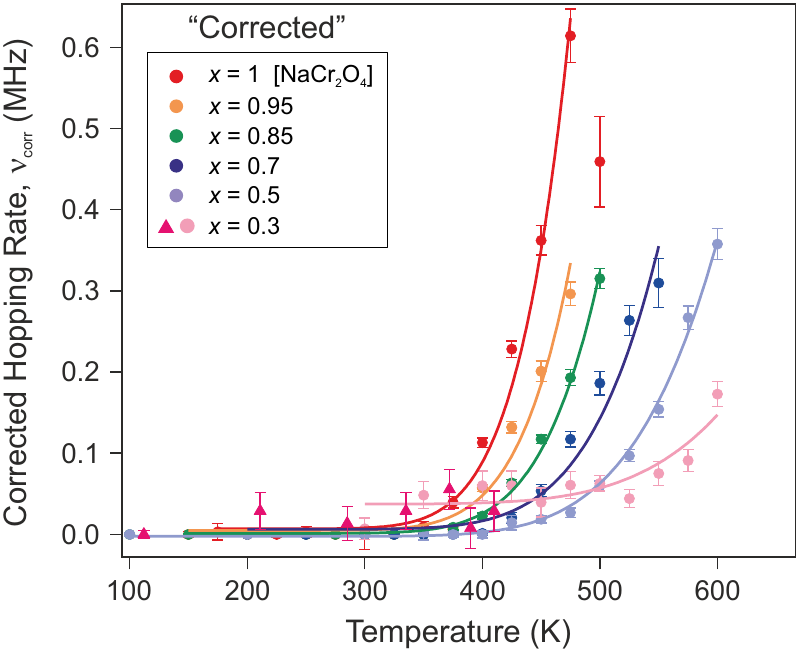}
  \end{center}
  \caption{Plot of the corrected hopping rate of the Na-ion diffusion process ($\nu_{\rm corr}$) as a function of temperature for several Na contents ($x$). The solid lines are fits to the data using the Arrhenius function (Eq.~\ref{arr}).}
  \label{hop2}
\end{figure}

Here a clarification is necessary, the G$_{\rm DGKT}$ relaxation function used in the fit originates from the strong collision model, according to which the dynamical behavior is described by a stochastic process. In particular, the local field at the muon site takes a certain value for a time, given by the inverse of the hopping rate, followed by a new value uncorrelated from the previous one. In order for a muon to detect the field fluctuation due to Na-ion diffusion within its lifetime, the amount of Na ions around the muons sites needs to be sufficient to maximize the detection probability. This means that the Na concentration will necessarily affect the field fluctuation rate felt by the muons, unless the Na diffusion length within the muon lifetime is long enough to cover more than one formula unit along the diffusion channel (e.g. 3 formula units in the $x=0.3$ case). If we assume that the Na diffusion length does not fulfill the latter condition, a correction factor is needed in order to account for the changes in the Na concentration. Intuitively, this factor will be proportional to the inverse of the Na concentration. Quantitatively, we may calculate the effect of the Na concentration on the local field in the Van Vleck limit:

\begin{equation}
\Delta^2_{ZF}=2(\frac{\mu_0}{4\pi})^2\sum_i\frac{\gamma^2_i \hbar^2}{r_i^6}\frac{I_i(I_i+1)}{3},
\label{correction}
\end{equation}
where $\Delta_{ZF}$ is the field distribution width at the muon sites; $\mu_0$ is the vacuum permeability; $\gamma_i$ is the gyromagnetic ratio of the i-th nucleus; $r_i$ is the distance between the i-th atom and the muon site; $I_i$ is the nuclear spin for the i-th atom. A plot of the measured $\sigma$ resulting from the fit to Eq.~\ref{3}, compared to the values calculated in Eq.~\ref{correction}, is displayed in Fig.~\ref{delta}. Since the nuclear moment of Cr and O is negligible, the Na nuclear moment is the sole contributor to the local field, which is therefore correlated to the number of Na contributing to the local field. Therefore, the correction factor is given by the ratios of the delta values with respect to the $x=1$ member of the solid solution. Table \ref{deltaval} lists all calculated values of $\Delta_{ZF}$ with the associated $\sigma$ and correction factors for each composition $x$. The resulting hopping rate is displayed in Fig.~\ref{hop2}.

\begin{table}[ht]
\caption{\label{deltaval}
Calculated values for $\Delta$, $\sigma$ and the correction factors $\Delta_{ratio}$ as a function of Na content ($x$). See text for more details.
}
\renewcommand{\arraystretch}{1.25}
\begin{tabular}{ccccc}
\hline
 Na content ($x$) &$\Delta$ [G]& $\sigma$ [1/$\mu$s] & $\Delta_{ratio}$ &\\
 \hline
 0.3 & 1.561900 & 0.133014 & 1.9138 &\\
 
 0.5 & 1.872519 & 0.159467 & 1.5964 & \\
 
 0.7 & 2.268891 & 0.193222 & 1.3175 &  \\
 
 0.85 & 2.685754 & 0.228723 & 1.1130 &\\
 
 0.95 & 2.886328 & 0.245804 & 1.0356 &  \\
 
 1 & 2.989218 & 0.254567 & 1 &\\
\hline
\end{tabular}
\end{table}

\begin{figure}[ht]
  \begin{center}
    \includegraphics[keepaspectratio=true,width=75 mm]{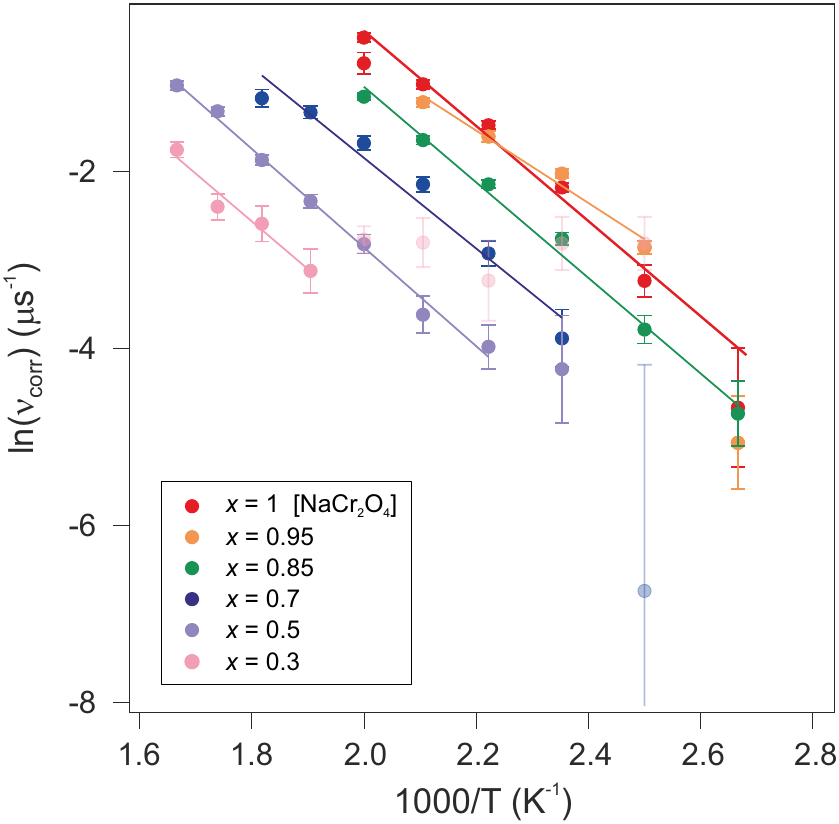}
  \end{center}
  \caption{Logarithmic hopping rate [ln($\nu_{\rm corr}$)] plotted as a function of inverse temperature for different Na content ($x$). The solid lines are linear fits to the data, yielding the activation energy $E_{\rm a}$.}
  \label{lnhop}
\end{figure}

We will refer to this corrected hopping rate ($\nu_{\rm corr}$) for the following discussion. The exponential increase of the hopping rate $\nu_{\rm corr}$ with temperature is related to the onset of Na-ion diffusion. Being this the case, the temperature dependence $\nu_{\rm corr}(t)$ display the signature of a thermally activated (diffusion) process, that can be well described by a simple model, the Arrhenius function \cite{Borg2012}:

\begin{eqnarray}
 \nu(T) = A \cdot e^{(-\frac{E_a}{k_BT})}.
\label{arr}
\end{eqnarray}
Here, $A$ is an empirical pre-exponential factor, with the dimensions of a frequency. This parameter accounts for the probability of an atom to make a diffusive jump \cite{Borg2012}. In the assumption of a Boltzmann-like energy distribution among the atoms in the system, the exponential term $exp(-\frac{E_a}{k_BT})$ represents the fraction of atoms that possess enough kinetic energy to overcome the energy barrier between the initial (static) and final (dynamic) state. Such energy barrier is the activation energy $E_{\rm a}$ in the exponential term of eq.~\ref{arr}, while $k_B$ is the Boltzmann constant ($8.62\cdot10^{-5}$~eV$\cdot$K$^{-1}$) and $T$ is the temperature in K. It is evident from the plot in Fig.~\ref{hop} that an increase in the Na content for the solid solution, causes a reduction in the temperature required to activate the process. Therefore, it might seem that the increase in the Na content induces a lowering in $E_{\rm a}$ required to start the diffusion. However, this is not really the case. In order to demonstrate this fact let us consider a different way to represent the hopping rate by plotting its logarithm as a function of the inverse temperature (see Fig.~\ref{lnhop}).

\begin{figure}[ht]
  \begin{center}
    \includegraphics[keepaspectratio=true,width=75 mm]{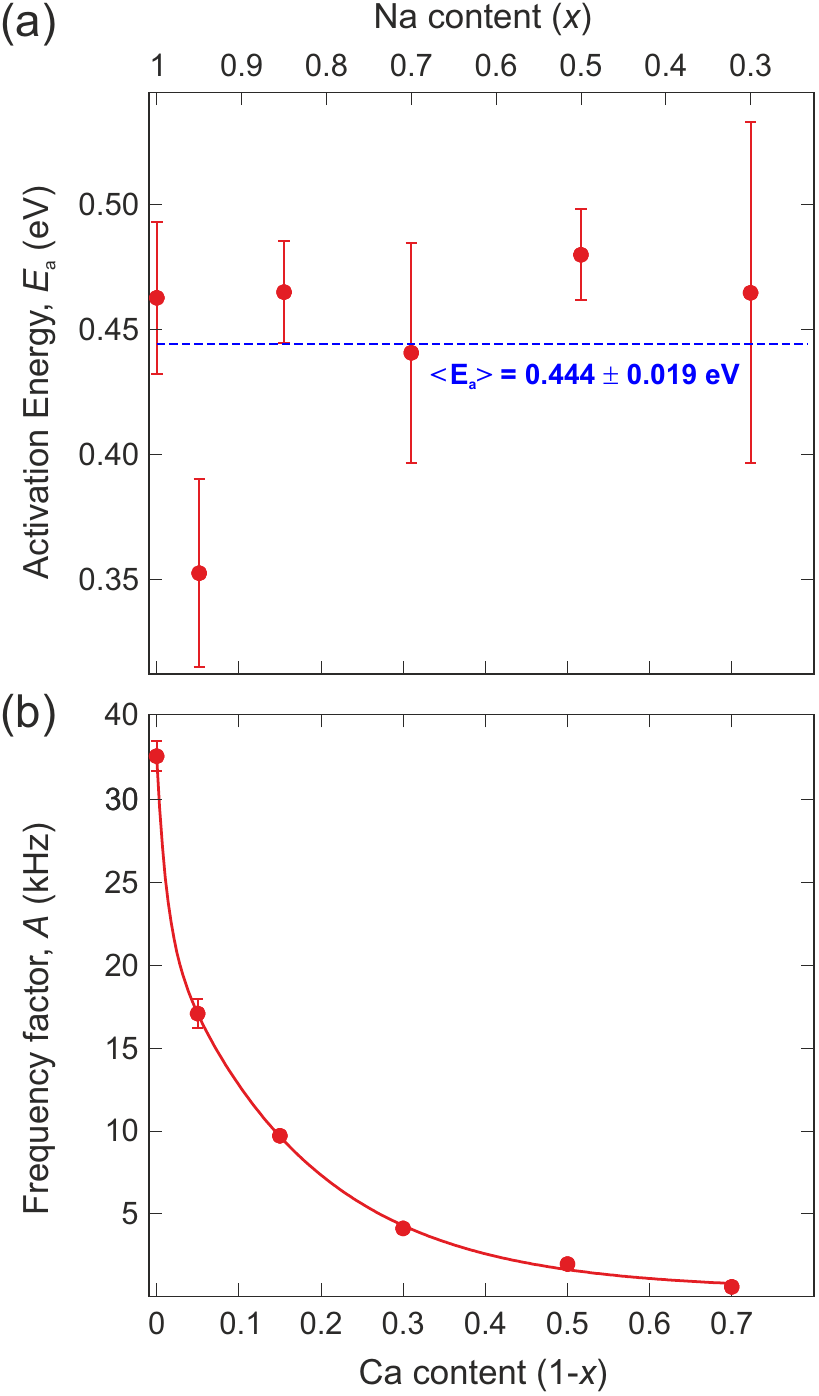}
  \end{center}
  \caption{(a) Values of the activation energy ($E_{\rm a}$) extracted from the logarithmic hopping rate (Fig.~\ref{lnhop}), the dashed line is a linear fit to the data providing the average value of the activation energy, $<E_{\rm a}>$. (b) Values of the pre-exponential factor ($A$) of the Arrhenius equation (Eq.~\ref{arr}) obtained from the fits in Fig.~\ref{hop2} and by using the average value of the activation energy $<E_{\rm a}>$. The continuous line is a fit to the data with an exponential decay function.}
  \label{Ea_A}
\end{figure}

Taking the logarithm on both sides of the Arrhenius Eq.~\ref{arr} it is easy to see that the values of the activation energy $E_{\rm a}$ for each sample can be determined via a linear fit (Fig.~\ref{lnhop}). The results of the linear fits are plotted in Fig.~\ref{Ea_A}(a), as a function of the Ca content. It is evident that $E_{\rm a}$ do not show an increasing trend as the Na content decreases, but oscillate slightly around an average value $\langle E_{\rm a} \rangle =(0.4605\pm0.0153)$~eV. The value of $E{\rm _a}$ in Eq.~\ref{arr} was then fixed to the average value $\langle E_{\rm a} \rangle $ and only the pre-factor $A$ was left free as a fitting parameter for the curves in Fig.~\ref{hop2}. The results of such fits are displayed in Fig.~\ref{Ea_A}(b).

The pre-exponential term $A$, is often neglected when discussing ion diffusion since it's difficult to experimentally determine it with confidence. In the present study, however, the very systematic approach and high quality of the data allowed us to robustly extract $A$ for each sample. The value of this parameter undergoes an exponential drop as the Na content decreases (i.e., substituted by Ca). This fact implies that in the system Na$_{\rm x}$Ca$_{1- \rm x}$Cr$_2$O$_4$ the replacement of Na with Ca has the effect of reducing the probability for a Na-ion to gain enough kinetic energy to perform a diffusive jump without modifying the potential energy landscape. As a result, the Na-ion mobility is systematically reduced and a higher temperature is required to activate the diffusion process. However, note that the activation energy ($E_{\rm a}$) necessary for the ions to overcome the potential barrier between the static state and the dynamic state remains more or less constant. The clear correlation between the Na content ad the systematic changes in the trend of $\nu$ allows to exclude the presence of a possible contribution from muon motion to the signal.

Finally, we would like to focus on the $x=0.95$ sample that seems to display a slightly lower activation energy than the other compositions [see Fig.~\ref{Ea_A}(a)]. Such effect could simply be related to fluctuations of the fitting routine. However, it is interesting that a very small amount of Ca 'defects' could in fact enhance the Na-ion diffusion. Such effect has previously been shown for Na-ion battery materials, where a surprisingly strong improvement was found for tiny atomic substitutions within the lattice \cite{Ma_2021}. Future detailed studies of the Na$_{\rm x}$Ca$_{1- \rm x}$Cr$_2$O$_4$ family for the range $x=1-0.9$, will be necessary to conclude if this is indeed a real effect.

\subsection{\label{diff1}Na-ion Diffusion Path}

\begin{figure*}[ht]
  \begin{center}
    \includegraphics[keepaspectratio=true,width=150 mm]{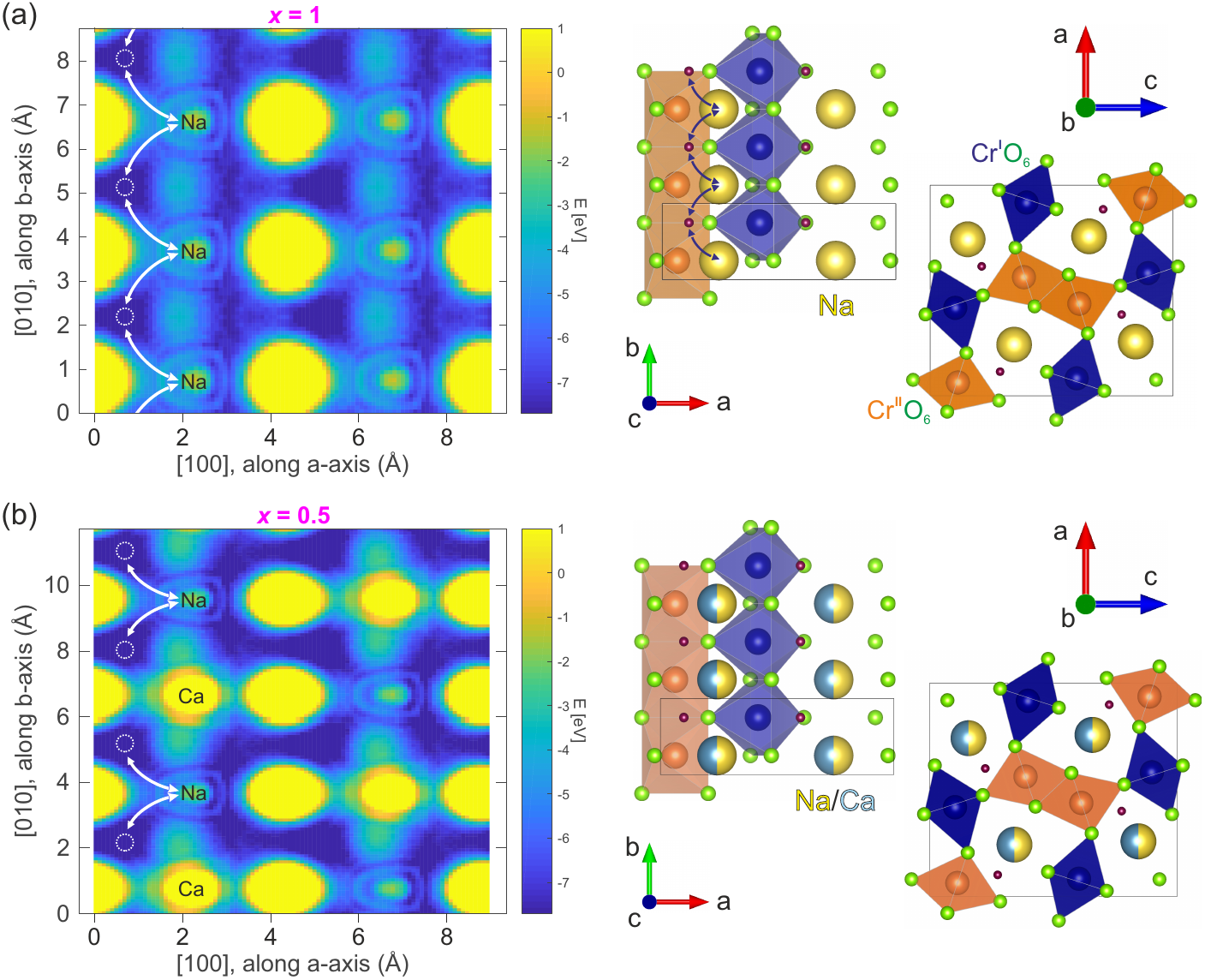}
  \end{center}
  \caption{(a) Left panel displayes the energy landscape of the system for $x=1$ within the Na-ion $ab-$plane. Right panel displays the corresponding crystal structure in two different orientations with the crystallographic site of the energy minima marked by burgundy red circles. (b) The same plots as in (a) for $x=0.5$ composition. In (a) and (b), three and two cells, respectively, were joined consecutively along the $b-$axis for clarity of display. Na and Ca sites are indicated and arrows schematically indicate the jump diffusion path to the interstitial site, which is marked by the dashed white circle in the left panels.}
 \label{Nacut}
\end{figure*}

In order to determine the diffusion coefficient, the diffusion path as well as the diffusion mechanism for the considered compounds need to be determined. The most probable diffusion mechanism taking place in Na$_{\rm x}$Ca$_{1- \rm x}$Cr$_2$O$_4$ is the interstitial \cite{Borg2012} jumping mechanism, since the Na/Ca Wycoff site 4c is fully occupied. Moreover, we will assume that the Ca is stationary and that Na is the only mobile species in this material family. Finally, only nominal~$\rightarrow$~interstitial~$\rightarrow$~nominal ~jumps are assumed to take place, while direct interstitial~$\rightarrow$~interstitial jumps are not considered (see also Fig.~\ref{Nacut}).

The detailed path of the interstitial mechanism is estimated based on the charge densities in the compound. This method has been utilized to estimate the diffusion path also in past studies on interstitial mechanism \cite{Sugiyama2011}. The charge density is calculated in the density functional theory (DFT) framework using the software package QUantum espresso \cite{QE-2009, QE-2017}. A simple self consistent calculation using the pseudo-potentials described by Refs.~\cite{Lejaeghere2016, Prandini2018}, result into charge densities whose associated electrostatic potential distribution is shown in Fig.~\ref{Nacut}(a,b) for $x=1$ and $x=0.5$, respectively. The plots display a cut along the $ab-$plane and the values of the energy, expressed in eV, are reported in the corresponding colorbar. The potential minima are characterized by a dark blue color in the figure. These results may be generalized for the other $x$ contents by asserting that the interstitial sites are barely affected by the presence of Ca. This is a reasonable assumption, considering that the crystal symmetry is weakly affected by the Ca doping (as clearly shown above by our NPD data).

In order to perform the ground state calculations of the $x=0.5$ system, Na and Ca ions in the provided crystallographic model have been placed in alternate sites along the $b-$axis, hereby doubling the unit cell. The corresponding non-doubled unit cell represented on the right side of Fig.~\ref{Nacut}(b) displays each Na/Ca site along the $b-$axis as having equal probability of being occupied by either Na or Ca atoms. The crystallographic positions of the interstitial jump sites for the $x=1$ and $x=0.5$ samples in fractional coordinates are [0.114(6), 0.750(1), 0.287(9)] and [0.100(2), 0.750(0), 0.275(0)], respectively.

\subsection{\label{diff2}Na-ion Diffusion Coefficient}

Since both the diffusion path and the diffusion mechanism have been determined, the diffusion coefficient for the title compounds can be calculated.
The diffusion coefficient is a concept based on the first and second law of Fick. The coefficient in itself is a macroscopic parameter and describes ultimately the flow of particles. The interesting microscopic details can, however, be derived based on the random walk approach which is, in many text books \cite{Borg2012}, described as:
\begin{eqnarray}
D_{\rm un}=\frac{nr^2}{2t},
\label{eq:D1}
\end{eqnarray}
where $nr^2$ is the total travel distance of the species; $t$ is the time required to travel such distance ($t=\frac{n}{\nu}$); $n$ is the average number of jumps. This description has in the past shown to be accurate to describe self diffusion of both Na and Li \cite{Sugiyama2009, Sugiyama2011, Forslund2020}. However, a term often neglected in past studies is the correlation factor ($f$). In general, each hop of the diffusing species is in fact correlated to the previous jump. Given that $f$ is simply a constant and that such values is somewhat close to 1, calculation of $f$ is often neglected and truthfully perhaps not necessary. However, $f$ highly depends on the crystalline environment of the diffusing species. In other words, there will be a systematic dependence on $f$ in this doping-dependent study. It is thus highly desirable to determine $f$ as a function of $x$ in order to calculate the Diffusion coefficient of Na$_{\rm x}$Ca$_{1- \rm x}$Cr$_2$O$_4$.
The correlation factor is given by the ratio of the diffusion coefficients of the real and the ideal uncorrelated one:
\begin{eqnarray}
f=lim_{n\rightarrow \infty} \frac{D_{\rm real}}{D_{\rm un}},
\label{eq:f}
\end{eqnarray}
where $D_{\rm un}$ is the uncorrelated diffusion coefficient given in Eq.~\ref{eq:D1}, whereas $D_{\rm real}$ is the actual diffusion coefficient of the system. Using Eq.~\ref{eq:D1} and the definition of diffusion coefficient in a none-random walk, the expression of $f$ is determined as follows:
\begin{eqnarray}
f=lim_{n\rightarrow \infty} 1+\frac{2}{n}\sum^{n}_j\sum^{n-j}_i \langle cos(\theta_{i,i+j}) \rangle,
\label{eq:f_1}
\end{eqnarray}
where $\theta_{i,i+j}$ is the angle between the ith and ith jump whereas $\langle \rangle$ denotes the average. Moreover, in an interstitialcy mechanism, $\langle cos(\theta_{i,i+j}) \rangle =0$ for $j\neq1$, such that Eq.~\ref{eq:f_1} is simplified into \cite{Compaan1956}
\begin{eqnarray}
f=lim_{n\rightarrow \infty} 1+\frac{1}{n}\sum^{n-1}_{i=1} \langle cos(\theta_{i,i+1}) \rangle.
\label{eq:f_2}
\end{eqnarray}

The above expression is valid for any diffusing species based on the interstitialcy mechanism. However, the extent of the summation changes depending on the detailed structure. From the considered diffusion path, $n$ varies as a function of $x$ given that the diffusion itself is hindered by the crystalline sites of the Ca. Naturally, $n$ is infinite in the $x=1$ case and Eq. \ref{eq:f_2} can be analytically solved into 1+$cos(\theta)$. For the other compositions $x$, $n$ is the number of jumps in the unique path that results into a contribution to $D$. Given the assumption that the Ca ions are uniformly distributed, the values of $n$ can be graphically estimated by simulating the jumping path of the Na ions between two consecutive Ca ions. Following this argument, a relation which describes the variation of $n$ as a function of the Ca concentration $\rho_{\rm Ca}$ can be extracted:
\begin{eqnarray}
n(\rho_{\rm Ca})=(\frac{1}{\rho_{\rm Ca}}-1) \cdot 2-1.
\label{eq:n}
\end{eqnarray}
The values of $n$ for each sample are summarised in Table~\ref{table:table1}, together with the angles $\theta_{i,i+1}$. The angles for the $x=1$ and $x=0.5$ samples are estimated from the calculated interstitial sites (see previous section). The angles for the other samples are extrapolated assuming a linear trend of the angle as a function of the composition. This is a reasonable assumption, given the linearity of the relation between the Na/Ca sites distances and the sample composition (see also Fig.~\ref{chan}). The correlation factor $f$ has been uniquely determined by using Eq.~\ref{eq:n} and Eq. \ref{eq:f_2}. The calculated $f$ for each considered concentration are listed in Table~\ref{table:table1}. With $f$ determined, the real diffusion coefficient can be calculated using Eq.~\ref{eq:D1} and Eq.~\ref{eq:f}.
\begin{table}[ht]
\caption{\label{table:table1}
The different values of $n$, $\theta$ and $f$ as a function of Na content ($x$). See text and Eqs.~\ref{eq:D1}-\ref{eq:n} for more details.}

\renewcommand{\arraystretch}{1.25}
\begin{tabular}{cccccc}
\hline
 Na content ($x$) &$\rho_{\rm Ca}$ &n & $\theta_{i,i+1}$&f&\\
 \hline
 0.3 & 3/2 &1& $87.217(2)^{\circ}$&1&\\
 
 0.5 & 1/2 &1& $89.810(4)^{\circ}$&1& \\
 
 0.7 & 1/3 &3&$92.402(8)^{\circ}$&0.972(1)&  \\
 
 0.85 & 1/7 &11&  $94.347(3)^{\circ}$&0.931(8)&\\
 
 0.95 & 1/20 &37&$95.643(7)^{\circ}$&0.904(3)&  \\
 
 1 & 0 &$\infty$& $96.292(3)^{\circ}$&0.890(4)&\\
\hline
\end{tabular}
\end{table}

\begin{figure}[ht]
  \begin{center}
    \includegraphics[keepaspectratio=true,width=77 mm]{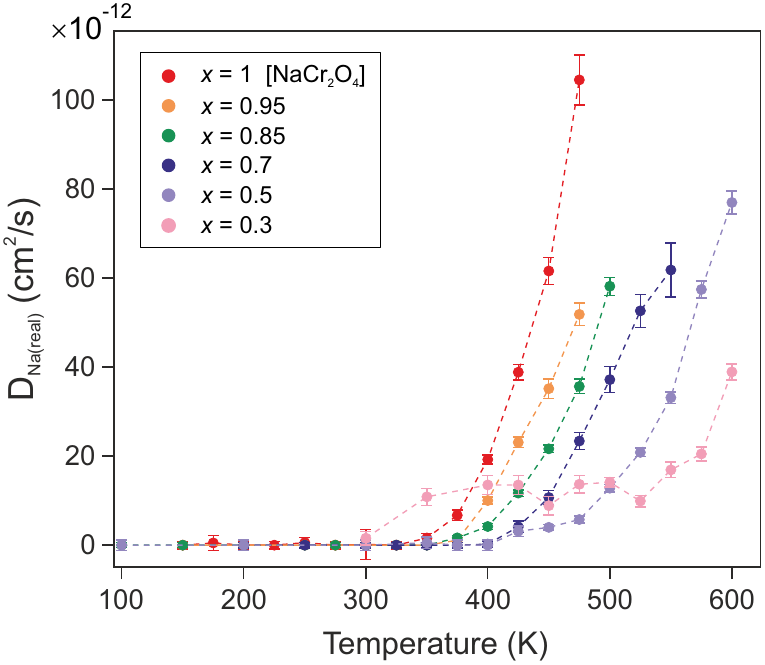}
  \end{center}
  \caption{Temperature dependence of the correlated diffusion coefficient $D_{\rm Na(real)}$ for all the members of the solid solution Na$_{\rm x}$Ca$_{1- \rm x}$Cr$_2$O$_4$.}
 \label{DNa_corr}
\end{figure}

The value $n=1$, calculated using the relation in Eq.~\ref{eq:n} for the $x=0.5$ composition, is considered as a boundary value adopted also for cases $x\leq0.5$. This choice is based on the fact that a negative value for $n$ would imply a negative number of jumps, which is un-physical.
The uncorrelated diffusion coefficient for the Na ions $D_{\rm Na(un)}$ for the members of the solid solution is estimated in a similar fashion as in reference \cite{Sugiyama2011}. Starting from the assumption that the hopping rate $\nu$ is the actual jumping rate of Na ions among neighboring sites, and treating the interstitial sites as vacancies, the general expression for $D_{\rm Na(un)}$ is the following \cite {Borg2012}:
\begin{eqnarray}
D_{\rm Na(un)}=\sum^{n}_{i=1} \frac{1}{N_i} Z_{v,i} s^2_i \nu.
\label{eq:DNa}
\end{eqnarray}
Here, $N_i$ is the number of Na sites in the $i$:th path, equal to 2 in our case; $Z_{v,i}$ is the vacancy fraction, equal to 1 since the interstitial site is unoccupied; $s_i$ is the jump distance between the nominal Na site and the interstitial site. The jump distances for the concentrations $x=1$ and $x=0.5$ are obtained directly from the calculations exposed in section \ref{diff1}. For the other concentrations, $s$ has been extrapolated assuming a linear trend as a function of $x$. Finally, the correlated diffusion coefficient $D_{\rm Na(real)}$ has been estimated using Eq.~\ref{eq:f}. The resulting temperature dependence for all the members of the solid solution is plotted in Fig.~\ref{DNa_corr}. Table~\ref{table:table2} reports the values of the parameters used in Eq.~\ref{eq:DNa} to calculate the diffusion coefficient for each sample as well as the values of the correlated diffusion coefficient $D_{\rm Na(real)}$ for a temperature $T=475$~K. 

\begin{table}[ht]
\caption{\label{table:table2}
Parameters from Eq.~\ref{eq:DNa} and correlated diffusion coefficient $D_{\rm Na(real)}$ at $T=450$~K as a function of Na content ($x$).
}

\renewcommand{\arraystretch}{1.25}
\begin{tabular}{cccc}
\hline
 Na content ($x$) &$s [$Å$] $ &$D_{\rm Na(real)}$ ($T=475$~K) [cm$^2$/s]&\\
 \hline
 0.3 & 2.122(9) &1.4(2)e-12&\\
 
 0.5 & 2.075(1) &5.8(7)e-12&\\
 
 0.7 & 2.027(2) &2.3(1)e-11&\\
 
 0.85 & 1.991(4) &3.6(2)e-11&\\
 
 0.95 & 1.967(4) &5.2(2)e-11&\\
 
 1 & 1.955(5) &1.04(5)e-10&\\
\hline
\end{tabular}
\end{table}

The exponential temperature dependence of the diffusion coefficient denotes the thermally activated nature of the process (Fig.~\ref{DNa_corr}). The value of the correlated diffusion coefficient for the pure Na compound ($x=1$) is one order of magnitude bigger than in the other compounds (see also Table~\ref{table:table2}). This behavior is to be ascribed to the concomitant contribution of several factors. The presence of Ca ions constitutes a physical impediment for the Na diffusion since it can only occur along the 1D channel. The mixed valence state Cr$^{3.5+}$ induced by the presence of Na implies an enhanced Cr-O bond stability, due to the lower occupation of $3d$ orbitals, and a reduction of the atomic radius of Cr atoms. As a consequence, a contraction of the transition metal oxide octahedra occurs, which results in a reduced volume for the 1D diffusion channels on one hand, and a weakened Na-O bonds on the other hand. The downsized 1D channels provide a more confined and advantageous diffusion path for the Na ions to move in a correlated fashion. The evolution of the electronic configuration throughout the solid solution members does not modify the energy barrier ($E_{\rm a}$), which hinders the Na diffusion. The observed enhancement of the Na mobility is therefore a purely geometrical effect.
As a final remark we might now calculate the upper value for the Na diffusion length $L_{\rm Na}$ within the muons mean lifetime $\tau=2.2~\mu$sec for the $x=1$ sample, to verify that our assumption was correct:
\begin{eqnarray}
L_{\rm Na(x=1)}=\sqrt{D_{\rm Na(real)}\tau} = 1.5 \text{Å}.
\label{eq:LNa}
\end{eqnarray}
Since the length of the unit cell along the direction of the 1D diffusion is $b$~=~2.9128(1) Å, considering a short diffusion length was a reasonable assumption. However, please note that we acquire and fit data up to $20~\mu$sec (see Fig.~\ref{muon}).

\section{\label{sec:discussion}Conclusions}
In this work a systematic doping-dependent study on the family Na$_{\rm x}$Ca$_{1- \rm x}$Cr$_2$O$_4$ with [$x=$~0.3,~0.5,~0.7,~0.85,~0.90,~0.95,~1], carried out using $\mu^+$SR and NPD methods is presented. The study shows how the Na kinetics can be tuned by means of the chemical pressure induced by the Ca doping. In particular, we observed that a reduced volume of the 1D diffusion channel corresponds to an enhancement of the Na ion mobility, contrary to the phenomenology of Li ion 1D battery materials. Moreover, the Ca-doping has the effect of reducing the probability for a Na ion to gain enough kinetic energy to overcome the activation energy barrier between the static and the dynamic state of the system without modifying the potential energy landscape. The ion diffusion process in Na$_{\rm x}$Ca$_{1- \rm x}$Cr$_2$O$_4$ is found to be an interstitial mechanism with highly correlated jumps. The diffusion coefficients for each member of the solid solution have been calculated taking into account the, usually neglected, correlation coefficient.

\section*{Acknowledgements}
The authors wish to thank J-PARC and ISIS/RAL for the allocated muon/neutron beam-time as well as their technical staff for the valuable help and great support during the experiments.
This research is funded by the Swedish Foundation for Strategic Research (SSF) within the Swedish national graduate school in neutron scattering (SwedNess), as well as the Swedish Research Council VR (Dnr. 2021-06157 and Dnr. 2017-05078), and the Carl Tryggers Foundation for Scientific Research (CTS-18:272). J.S. is supported by the Japan Society for the Promotion Science (JSPS) KAKENHI Grant No. JP18H01863 and JP20K21149. Y.S. and O.K.F. are funded by the Chalmers Area of Advance - Materials Science.



\balance


\bibliography{refs} 
\bibliographystyle{rsc} 

\end{document}